\documentclass[a4paper, 12pt, oneside]{article}
\usepackage[cp1251]{inputenc}
\usepackage[english]{babel}
\usepackage{graphics, amsfonts, amsmath, bm, amssymb, array}
\usepackage[dvips]{graphicx}
\usepackage[flushleft,small]{caption2}

\newcommand{\diag}{\rm \diag\, }
\newcommand{\intl}{\int\limits}

\renewcommand{\Re}{\mathop{\rm Re\,}}
\renewcommand{\Im}{\mathop{\rm Im\,}}

\makeatother {

\begin{document}
\thispagestyle{empty} \large

\renewcommand{\refname}{\begin{center}{\bf REFERENCES}\end{center}}

\begin{center}
{\bf Surface Plasmons in Thin Metallic Films for the Case of
Antisymmetrical Configuration of Magnetic Field}
\end{center}

\begin{center}
  \bf  A. V. Latyshev\footnote{$avlatyshev@mail.ru$} and
  A. A. Yushkanov\footnote{$yushkanov@inbox.ru$}
\end{center}\medskip

\begin{center}
{\it Faculty of Physics and Mathematics,\\ Moscow State Regional
University,  105005,\\ Moscow, Radio st., 10--A}
\end{center}\medskip

\begin{abstract}
For the first time it is shown that for thin metallic films
thickness of which does not exceed thickness of skin -- layer, the
problem of description of surface plasma oscillations allows
analytical solution under arbitrary ratio of length of electron mean
free path and thickness of a film. The dependance of frequency of
surface plasma oscillations on wave number is deduced. We consider a
case of specular -- diffusive boundary value problems.
\medskip

\medskip

\end{abstract}

\begin{center}\bf
  Introduction
\end{center}

Electromagnetic properties of metal films has been being a subject
of great interest for a long time already \cite{F66} -- \cite{F09}.
Problem of surface plasma oscillations has been a problem of special
interest in recent time \cite{Economou} -- \cite{Ina}. It is
connected as with theoretical interest to this problem, and with
numerous practical appendices as well. Thus the majority of
researches is founded on the description of properties of films with
use of methods of macroscopical electrodynamics. Such approach is
inadequate for thin films, since macroscopical electrodynamics is
inapplicable for the description of films in the thickness of an
order of length of mean free path of electrons and less than this
length. The electrons scattering on a surface demands kinetic
consideration. It complicates the problem significantly.

In the present work it is shown that for thin films, a thickness of
which does not exceed a thickness of a skin -- layer, the problem of
description of surface plasma oscillations allows analytical
solution under arbitrary ratio between length of mean free path of
electrons and thickness of a film. The given work is a continuation
of our work \cite{LYarx2010} in which the case when values of a
magnetic field on top and bottom surface of films coincide was
considered . Now the situation, when the signs of these values
differ is investigated.

Let us note, that the most part of reasonings carryied out below is
true for more general case of conductive  (in particular,
semi-conductor) film.\medskip \vskip5mm

\begin{center}
{\bf Problem statement}
\end{center}

Let us consider a thin metal film.

We take Cartesian coordinate system with origin of coordinates on
one of the surfaces of a slab, with axis $x$, directed deep into the
slab and perpendicularly to the surface of the film. The axis $z$ we
will direct along the direction of propagation of the surface
electromagnetic wave. We will note, that in this case magnetic field
is directed along the axis $y$.

Under such choice of system of coordinates the electric field vector
and magnetic field vector have the following structure
$$
\mathbf{E}
=\{E_x(x,z,t),0,E_z(x,z,t)\}, \quad
\mathbf{H}
=\{0,H_y(x,z,t),0\}.
$$

The origin of coordinates we will place at the bottom plane limiting
a film. Let us designate a thickness of the film through $d$.

Out of the film the electromagnetic field is described by the equations
$$
\dfrac{1}{c^2}\dfrac{\partial^2 {\bf E}}{\partial t^2}-
\Delta {\bf E}=0
$$
and
$$
\dfrac{1}{c^2}\dfrac{\partial^2 {\bf H}}{\partial t^2}-\Delta{\bf
H}=0.
$$

Here $c$ is the velocity of light, $\Delta$ is the Laplace operator.

The solution of these equations decreasing at infinity point has the
following form
$$
{\bf E}=\left\{\begin{array}{l}{\bf E}_1e^{-i\omega t+\alpha x+ikz},\qquad
x<0, \\
{\bf E}_2e^{-i\omega t+\alpha (d-x)+ikz},\qquad x>d,
\end{array}\right.
\eqno{(1a)}
$$
and
$$
{\bf H}=\left\{\begin{array}{l}{\bf H}_1e^{-i\omega t+\alpha x+ikz},\qquad
x<0, \\
{\bf H}_2e^{-i\omega t+\alpha(d-x)+ikz},\qquad x>d.
\end{array}\right.
\eqno{(1b)}
$$

Here $\omega$ is the wave frequency, $k$ is the wave number, damping
parameter $\alpha$ is connected with these quantities by relation
$$
\alpha=\sqrt{k^2-\dfrac{\omega^2}{c^2}},
\eqno{(2)}
$$
$\mathbf{E}_j$ and $\mathbf{H}_j \;(j=1,2)$ are constant amplitudies.

Further we search components of intensity vectors of electric and
magnetic fields in the following form
$$
E_x(x,z,t)=E_x(x)e^{-i\omega t+ikz},\quad
E_z(x,z,t)=E_z(x)e^{-i\omega t+ikz},
$$
and
$$
H_y(x,z,t)=H_y(x)e^{-i\omega t+ikz}.
$$

Then the behaviour of electric and magnetic fields of the wave
within the film is described by the following system of the
differential equations (\cite{K} and \cite{Landau8})
$$
\left\{\begin{array}{l}
\dfrac{dE_z}{dx}-ik E_x+\dfrac{i\omega}{c}H_y=0, \\ \\
\dfrac{i\omega}{c}E_x-ik H_y=\dfrac{4\pi}{c}j_x,\\ \\
\dfrac{dH_y}{dx}+\dfrac{i\omega}{c}E_z=\dfrac{4\pi}{c}j_z.
\end{array}\right.
\eqno{(3)}
$$

Here  $\,\bf j$ is the current density.

The equations (3) are satisfied out of the slab under the condition
${\bf j}=0$ as well.

Then impedance at the bottom surface of the layer (film) is defined
as follows
$$
Z^{(j)}=\dfrac{E_z(-0)}{H_y(-0)}, \qquad j=1,2.
\eqno{(4)}
$$

Quantities $Z^{(j)} \; (j=1,2)$ correspond to an impedance on the
bottom layer surface. At the same time the quantity $Z^{(1)}$
corresponds to magnetic field--symmetrical configuration of an
external field. This is the case $j=1$, when
$$
H_y (0) =H_y (d), \qquad E_x (0) =E_x (d), \qquad E_z (0) =-E_z (d).
\eqno {(5)}
$$

This case has been considered in \cite {LYarx2010}.

The quntity $Z^{(2)} $ corresponds to configuration of an external
field antisymmetric by magnetic field. It is the case $j=2$, when
$$
H_y (0) =-H_y (d), \qquad E_x (0) =-E_x (d), \qquad E_z (0) =E_z (d).
\eqno {(6)}
$$

It is required to find a spatial dispersion of the surface plasmon,
i.e. to find dependence of frequency of oscillations of eigen mode
of the system (3) on quantity of the wave vector $\omega
=\omega(k)$.

Let us consider a case when the width of the layer $d$ is less than
the depth of the skin -- layer $\delta$. We will note, that the
depth of a skin -- layer  depends essentially on frequency of
radiation, monotonously decreasing in the process of growth of the
last quantity. The quantity $\delta$ possesses minimum value in
so-called infra-red case \cite {Landau10} $ \delta_0 =\dfrac
{c}{\omega_p}, $ where $\omega_p$ is the plasma frequency.

For typical metals \cite{Landau10} $\delta_0\sim 10^{-5}$ cm.

Thus for the films which thickness $d $ is less $\delta_0$, our
assumption is true for any frequencies.

Quantities $H_y$ and $E_z$ vary a little on distances smaller than
the depth of a skin -- layer. Therefore at performance of the given
assumption ($d <\delta_0$)
thess fields will vary inside of slab. \\

\begin{center}
{\bf Surface plasmon. Antisymmetric configuration of magnetic field}
\end{center}

Let us consider the case 2 when $E_z(0)=E_z(d)$. We can assume that
in this layer $z$ -- projection of the electric field $E_z$ is
constant. Then magnetic field change on the width of a layer can be
defined from the third equation of the system (1)
$$
H_y(d)-H_y(0)=-\dfrac{i\omega}{c}E_zd+\dfrac{4\pi}{c}\intl_0^d
j_z(x)dx.
\eqno{(7)}
$$

Thus
$$
j_z(x)=\sigma(x)E_z,
$$
where $\sigma(x)$ is the conductivity which in general case
depends on coordinate  $x$.

We introduce conductivity averaged by the slab thickness
$$
\sigma_d=\dfrac{1}{E_zd}\intl_0^d j_z(x) dx=
\dfrac{1}{d}\intl_0^d \sigma(x) dx.
$$

Now we can rewrite the relation (7) in the form
$$
H_y(d)-H_y(0)=-\dfrac{i\omega}{c}E_zd+\dfrac{4\pi \sigma_d d }{c}E_z.
$$

Considering symmetry of a magnetic field, from here we have
$$
H_y(0)=\dfrac{i\omega d E_z}{2c}\Big(1+i\dfrac{4\pi\sigma_d}{\omega}\Big).
$$

Then for impedance (4) we have
$$
Z^{(2)}=-\dfrac{2ic}{\omega d\Big(1+i\dfrac{4\pi\sigma_d}{\omega}\Big) }.
\eqno{(8)}
$$

In the same way as well as in (9), we receive
$$
Z^{(2)}=\dfrac{i\alpha c}{\omega}.
\eqno{(9)}
$$

The dispersive equation for a surface plasma wave can be derived
from the expressions (2), (8) and (9)
$$
\dfrac{2c}{\omega d+4\pi i \sigma_d d }=
-\dfrac{ \sqrt{c^2k^2-\omega^2}}{\omega}.
 \eqno{(10)}
$$

From the dispersive equation (10) we find the required spatial
dispersion
$$
k(\omega)=\dfrac{\omega}{c}\sqrt{1+\dfrac{4c^2}{\omega^2d^2\Big(1+
\dfrac{4\pi i}{\omega}\sigma_d\Big)^2}}.
\eqno{(11)}
$$

Let us assume, that boundary conditions are specular -- diffusive,
$p $ is the specularity coefficient. Let the relation $kd\ll 1$ be
true. Then in a low-frequency case, when $\omega\to 0$, the quantity
$\sigma_d$ can be presented in the form \cite{S}
$$
\sigma_d=\dfrac{w}{\Phi(w)}\,\sigma_0,\quad\quad
w=\dfrac{d}{l},
\eqno{(12)}
$$
and
$$
\dfrac{1}{\Phi(w)}=\dfrac{1}{w}-\dfrac{3}{2w^2}(1-p)\intl_1^\infty\Big(
\dfrac{1}{t^3}-\dfrac{1}{t^5}\Big)\dfrac{1-e^{-wt}}{1-pe^{-wt}}dt.
\eqno{(13)}
$$

Here $l$ is the mean free path of electrons, $p$ is the coefficient
of specular reflection (specularity coefficient),
$\sigma_0=\omega_p^2\tau/(4\pi)$ is the static conductivity of
volume pattern, $\tau=l/v_F$ is the electron-transit time, $v_F$ is
the Fermi velocity.

For arbitrary frequencies thee expression (12) and (13) will hold
true under the condition, that it is necessary to use the following
expression $ l\to (v_F \tau)/(1-i\omega\tau) $ as quantity $l$, and
instead of $\sigma_0$ it's necessary to use the expression $
\sigma_0\to \sigma_0/(1-i\omega\tau). $

Let us reduce the formula (11) to the form convenient for numerical
calculations. We introduce the dimensionless parameters
$\varepsilon=\dfrac{\nu}{\omega_p}$ and
$\Omega=\dfrac{\omega}{\omega_p}$. Then we can transform the formula
(19) to the following form
$$
k(\Omega,\varepsilon)=\sqrt{\omega_p^2\dfrac{\Omega^2}{c^2}
+\dfrac{4}{d^2}
\Big(1-\dfrac{\varphi(w)}{\Omega(\Omega-i\varepsilon)}\Big)^{-2}},
\eqno{(14)}
$$
where
$$
\varphi(w)=1-\dfrac{1.5}{w}(1-p)\int\limits_{1}^{\infty}
\Big(\dfrac{1}{t^3}-\dfrac{1}{t^5}\Big)\dfrac{1-e^{-wt}}{1-pe^{-wt}}dt.
$$

In the case, when electrons reflect under specular condition from
the film surface (i.e. at $p=1$), the formula (14) becomes simpler
and looks like
$$
k(\omega)=\dfrac{\omega}{c}\sqrt{1+\dfrac{c^2(\nu-i\omega)^2}
{d_0^2(\nu\omega-i\omega^2+i\omega_p^2)^2}},
\eqno{(15)}
$$
where $d_0=\dfrac{d}{2}$ is the half of a film thickness.

In dimensionless parametres the formula (15) can be written as
follows
$$
k(\Omega)=\dfrac{\omega_p}{c}\Omega\sqrt{1+
\dfrac{c^2(\Omega+i\varepsilon)^2}
{(\omega_pd)^2(\Omega^2-1+i\varepsilon\Omega)^2}}.
\eqno{(16)}
$$

Under $\varepsilon=0$ we receive from here that
$$
k(\Omega)=\dfrac{\omega_p}{c}\Omega\sqrt{1+
\Big(\dfrac{c}{\omega_pd_0}\Big)^2\Big(\Omega-\dfrac{1}{\Omega}\Big)^{-2}}.
$$

It is clear that under $\varepsilon=0$ from the last formula it
follows that $\Im k(\Omega)=0$, i. e. in collisionless plasma
plasmon damping is absent.

From the formula (16) it is visible, that there exists such critical
frequency $\omega_0 =\omega_0 (\varepsilon, d)$, that under $\omega
<\omega_p $ $ \Im k(\omega)> \Re k(\omega)$, i.e. in field of
subcritical frequencies sufrace plasmons do not exist.

We adduce the table of critical frequencies, referring to plasma
(Langmuir) frequencies,
$\Omega_0(\varepsilon,d)=\dfrac{\omega_0(\varepsilon,d)}{\omega_p}$,
under $\varepsilon=10^{-1}$ for the case of specular boundary
conditions ($p=1$).

Table 1 (critical frequencies)

\begin{tabular}{|c|c|c|c|c|c|}
  \hline
Film thickness $d$, nm & & 1     & 2   & 3   & 4 \\ \hline
Critical frequency $\Omega_0$ &  & 0.101
&0.100&0.097&0.092\\ \hline \hline
Film thickness $d$, nm&5   & 6   &7   & 8 & 9\\ \hline
Critical frequency  $\Omega_0$ &0.086&0.078&0.067&0.051&0.023 \\
  \hline
\end{tabular}

Let us adduce graphics on Figs. 1 -- 8 of dependencies of real and
imaginary parts of the wave vector on the ratio of frequencies
$\omega/\omega_p $ under various values of frequency of electron
collisions $\nu$, thickness of a film $d$ and coefficient of
specular reflection $p$. We will consider the case of sodium films,
i.e. we take $\omega_p=6.5\cdot 10^{15}sec^{-1}$, $v_F=8.52\cdot
10^7$ cm/sec.

Depending on quantity of parametres $\nu, d, p$ the quantities
$\Re k$ and $\Im k$ can essentially differ. So, at
$\nu=10^{-5}\omega_p$, $d=10$ nanometer and $p=1$ (Fig. 1)
the quantity $\Re k$ surpasses $\Im k$ on some orders.

From Fig. 1 it is visible, that
if to enter quantity $Z =\dfrac{\Re k}{\Im k}$,  then
$Z(0.1,10^{-5},10,1)=2.1\cdot 10^{4}$, $Z(0.5,10^{-5}, 10,1) =3.8\cdot
10^{4}$.

Besides, the quantity $\Re k$ is always positive under all values of
parameters $ \omega, \nu, d, p $, while the quantity $\Im$ can be
negative in the field of superhigh frequencies as well.

Let us stop on existence of surface plasma waves
(see Figs. 2 and 3).

Depending on quantities $ \varepsilon, d, p $ two critical
frequencies $ \omega_0$ and $ \omega_1$, such, that the inequality $
\Im k <\Re k $ is true under $ \omega_0 <\omega <\omega_1$, and the
inequality $ \Im k> \Re k $ is true under $0 <\omega <\omega_p $ or
$ \omega_1 <\omega <\omega_p $ can exist.

In the last case of surface plasma waves do not exist. Let us
consider the case $ \nu=10^{-1} \omega_p, p=0.1$. Results
calculations we will present in the following table.

Table 2 (critical frequencies)

\begin{tabular}{|c|c|c|}
  \hline
  Film thickness, nm& First critical &  Second critical \\
         $d$         & frequency, $\omega_0$& frequency, $\omega_1$    \\
                     \hline
  1 & 0.168 & 0.904 \\
  2 & 0.130 & 0.924 \\
  3 & 0.116 & 0.929 \\
  4 & 0.107 & 0.932 \\
  5 & 0.098 & 0.934 \\
  6 & 0.089 & 0.935 \\
  7 & 0.077 & 0.935 \\
  8 & 0.063 & 0.936 \\
  9 & 0.041 & 0.936 \\
  10& 0.000 & 0.937 \\
  \hline
\end{tabular}

The behaviour of the real and imaginary parts of a wave vector in
dependence on a thickness of a film is presented on Figs. 4 and 5.
The real part has a sharp maximum nearby the plasma resonance $
\omega\sim \omega_p $.

Let us note, the more is thickness of a film the less are values of
the real part under each value of frequency of oscillations of an
electromagnetic field. The imaginary part has the same behaviour.
But its values are less than values of real parts significantly.
Unlike the real part in the field of the superhigh frequencies the
values of the imaginary part become negative.

On Figs. 6 and 7 dependences of the real and imaginary parts of wave
vector on quantity of collision frequencies of electrons are
presented. The less is the quantity $\nu$, the more are the values
of the real part. For the imaginary part inverse relation takes
place. Namely, the less is the quantity of collision frequencies,
the less is the value of an imaginary part. It means, that at
electron collisions frequency increase the attenuation of surface
plasma waves becomes stronger (in the field of subcritical
frequencies).

On Fig. 8 dependence of an imaginary part of the wave vector on
quantity of coefficient of specular reflection in the field of
subcritical frequencies is presented. Graphics show, that with the
growth of coefficient of specular reflection the values of an
imaginary part decrease. It means, that damping of plasma waves by
that becomes the stronger, the less are quantities of coefficient of
specular reflection.

\begin{center}
{\bf Conclusion}
\end{center}

In the present work the dispersion relation for surface plasmons is
deduced. We consider the case of an antisymmetric configuration of
$x$ -- component of the electric field and $y$ -- component of
magnetic field, and symmetric $z$ -- component of electric field. We
consider the case of specular -- diffusive boundary value
conditions.


\begin{figure}[b]\center
\includegraphics[width=16.0cm, height=9cm]{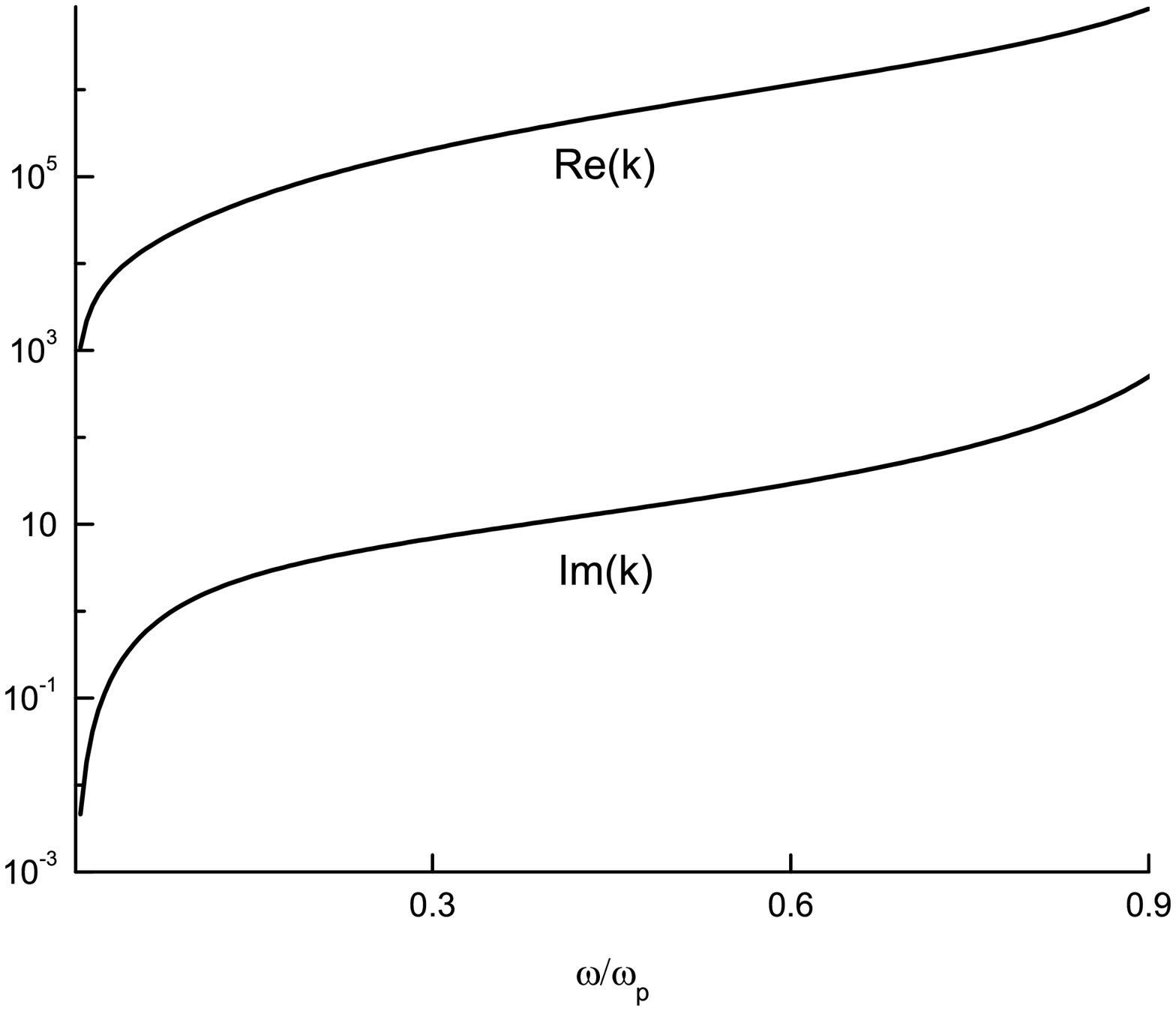}
\noindent\caption{Real and imaginary parts of wave number,
film thickness $d=10$ nm, collision electron frequency
$\nu=10^{-5}\omega_p$ 1/sec, specular reflection, $p=1$.
}\label{rateIII}
\includegraphics[width=16.0cm, height=9cm]{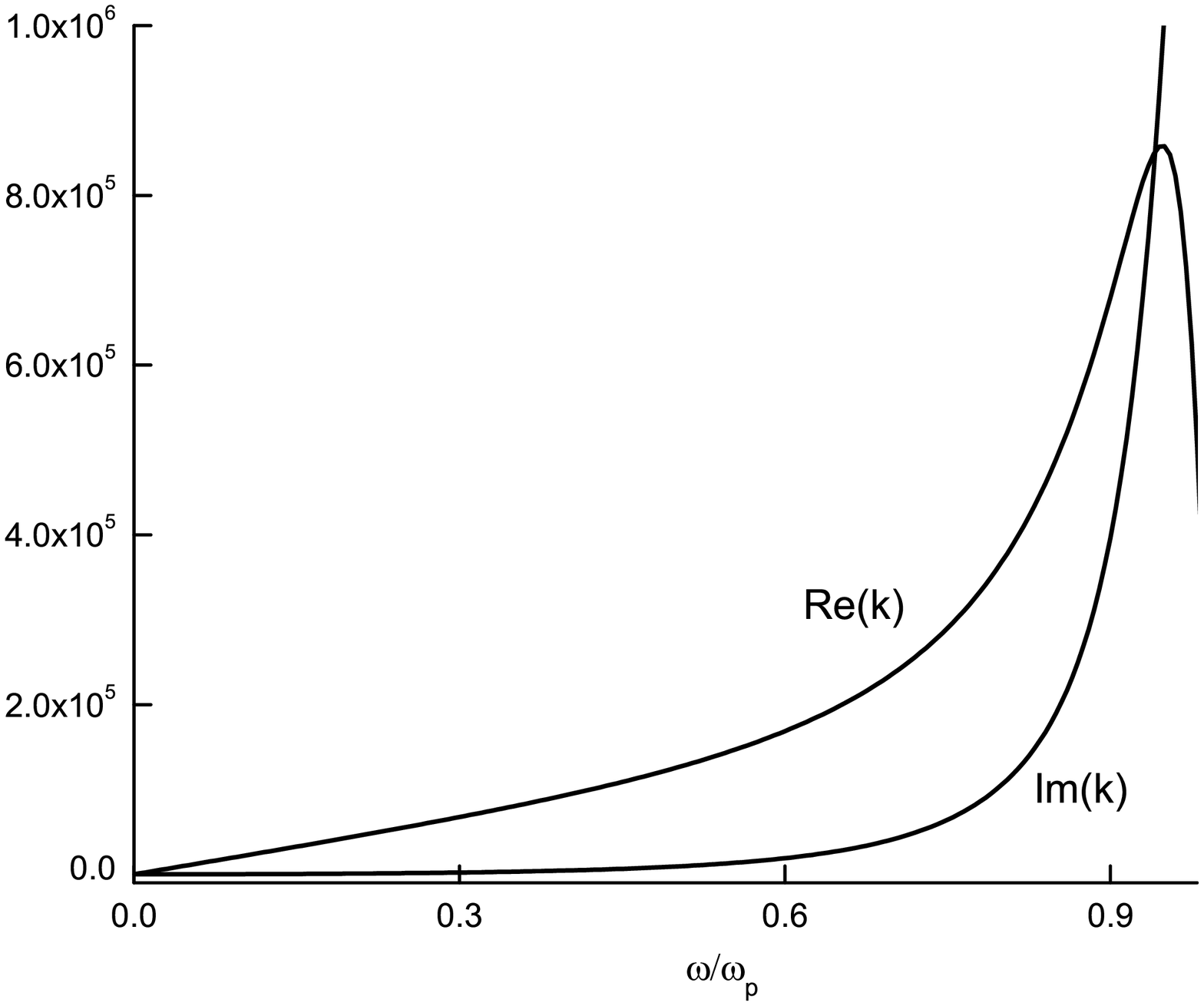}
\noindent\caption{Real and imaginary parts of wave number, film
thickness $d=100$ nm, collision electron frequency $\nu=0.1\omega_p$
1/sec, coefficient of specular reflection, $p=0.1$. }\label{rateIII}
\end{figure}

\begin{figure}[h]\center
\includegraphics[width=16.0cm, height=9cm]{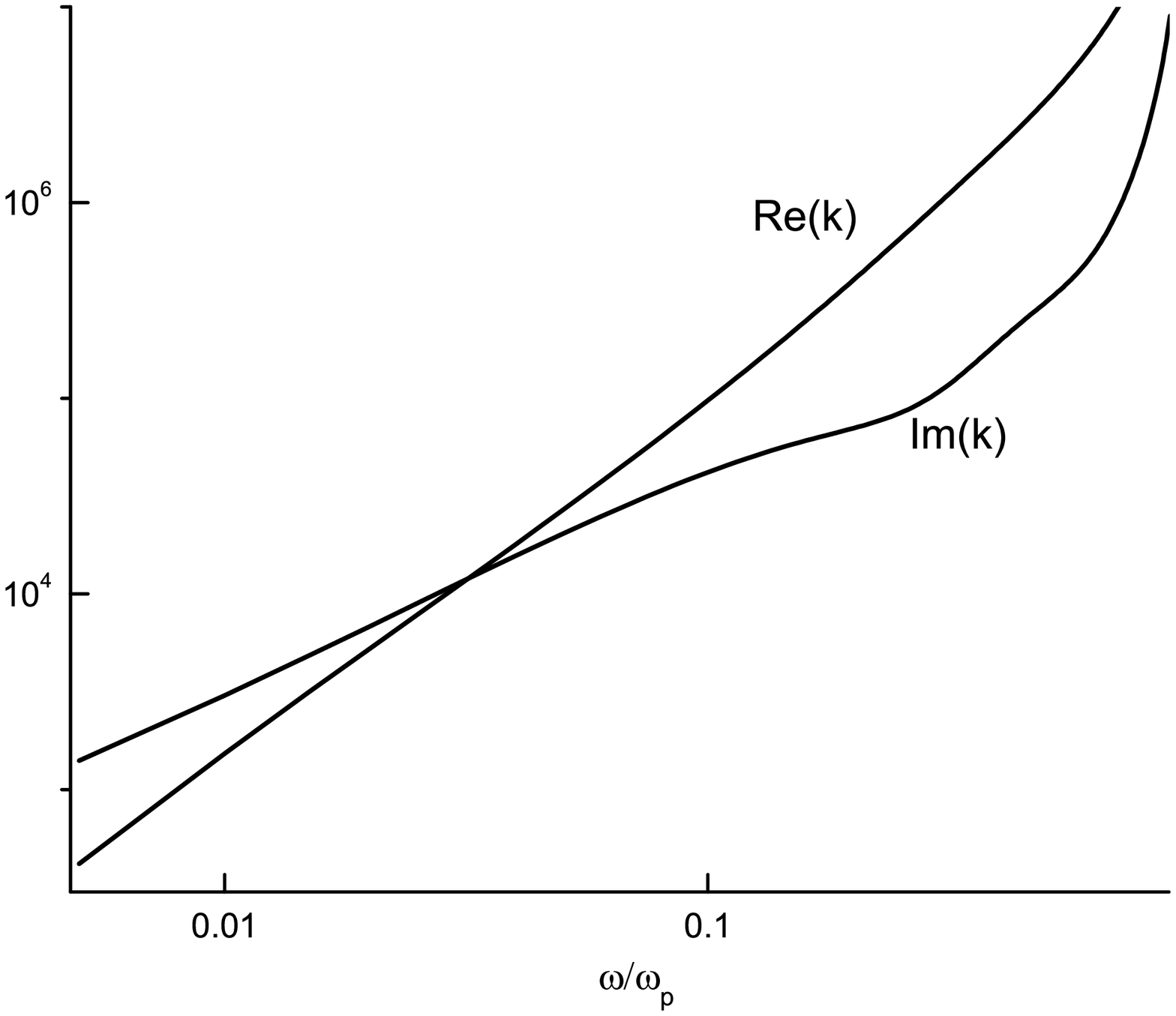}
\noindent\caption{Real and imaginary parts of wave number,
film thickness $d=100$ nm, collision electron frequency
$\nu=0.02\omega_p$ 1/sec, coefficient of specular reflection, $p=0.1$.
}\label{rateIII}
\includegraphics[width=16.0cm, height=9cm]{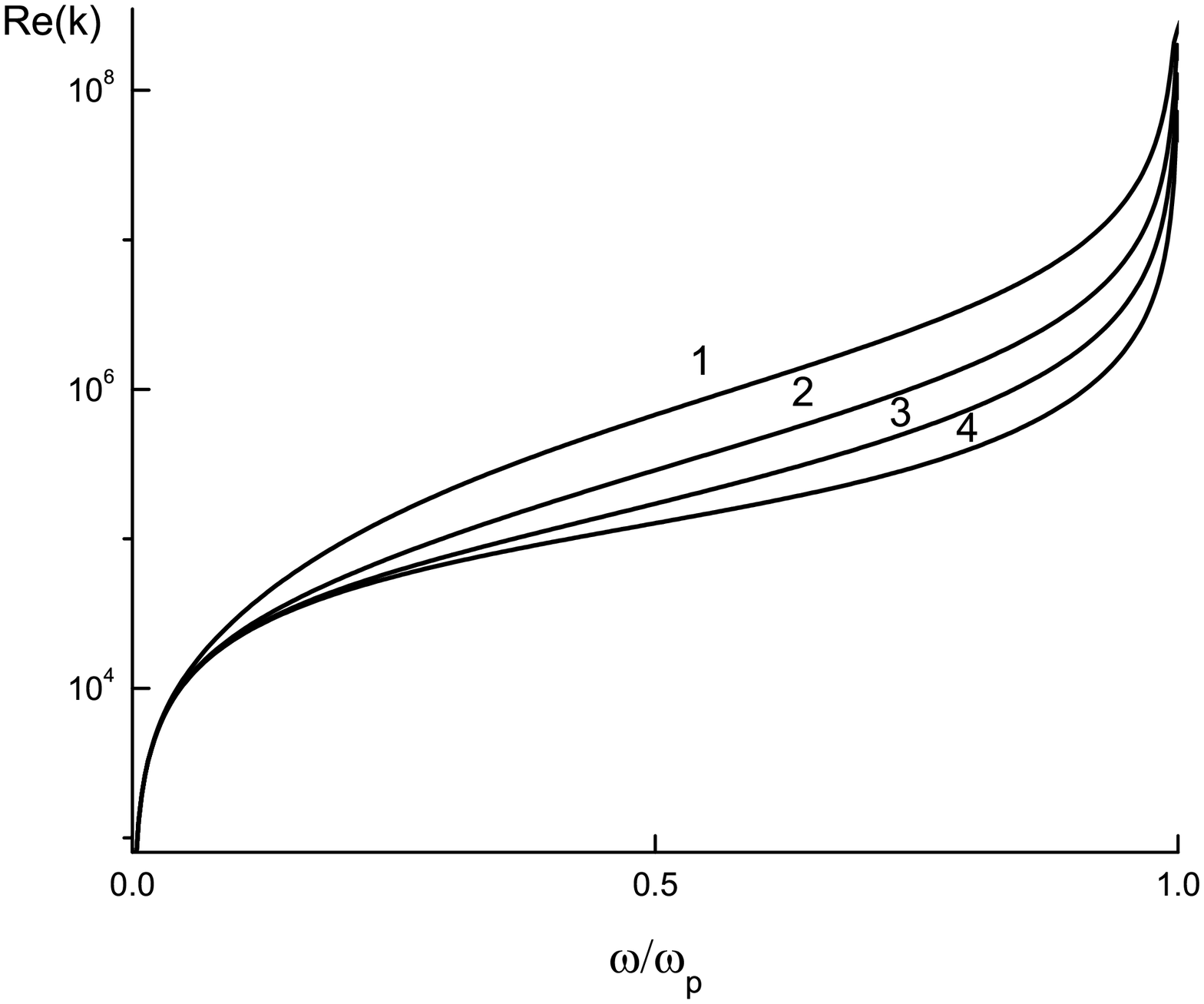}
\noindent\caption{Real and imaginary parts of wave number, collision
electron frequency $\nu=10^{-3}\omega_p$ 1/sec, $p=0.5$, curves
$1,2,3,4$ correspond to values of the film thickness
$d=10,25,50,100$ nm. }\label{rateIII}
\end{figure}

\begin{figure}[h]\center
\includegraphics[width=16.0cm, height=8cm]{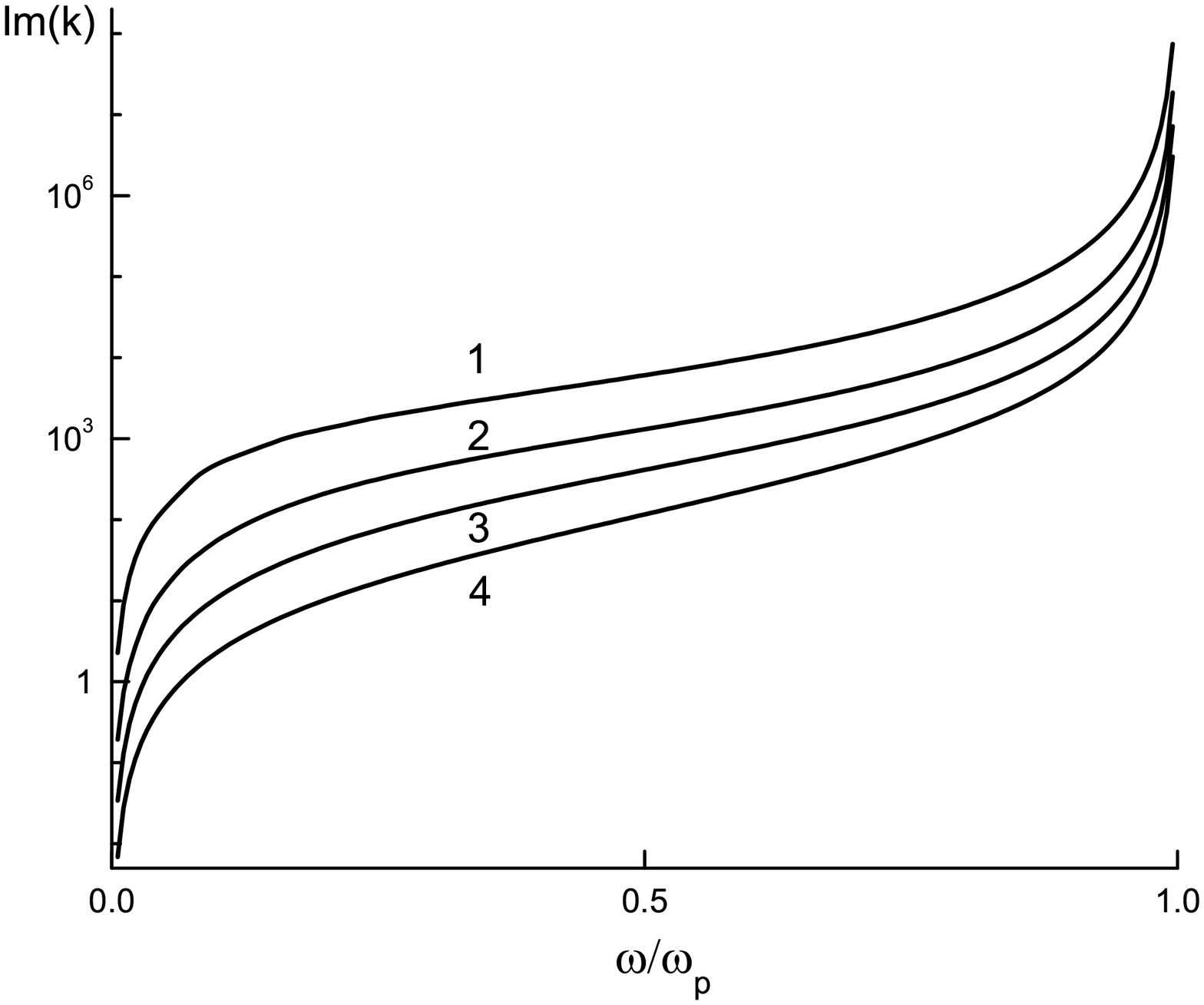}
\noindent\caption{Imaginary part of wave number,
$\nu=10^{-3}\omega_p$ 1/sec, $p=0.5$,
curves $1,2,3,4$ correspond to values of the film thickness
$d=10,25,50,100$ nm. Subcritical frequencies: $0<\omega<\omega_p$ 1/sec.
}\label{rateIII}
\includegraphics[width=16.0cm, height=8cm]{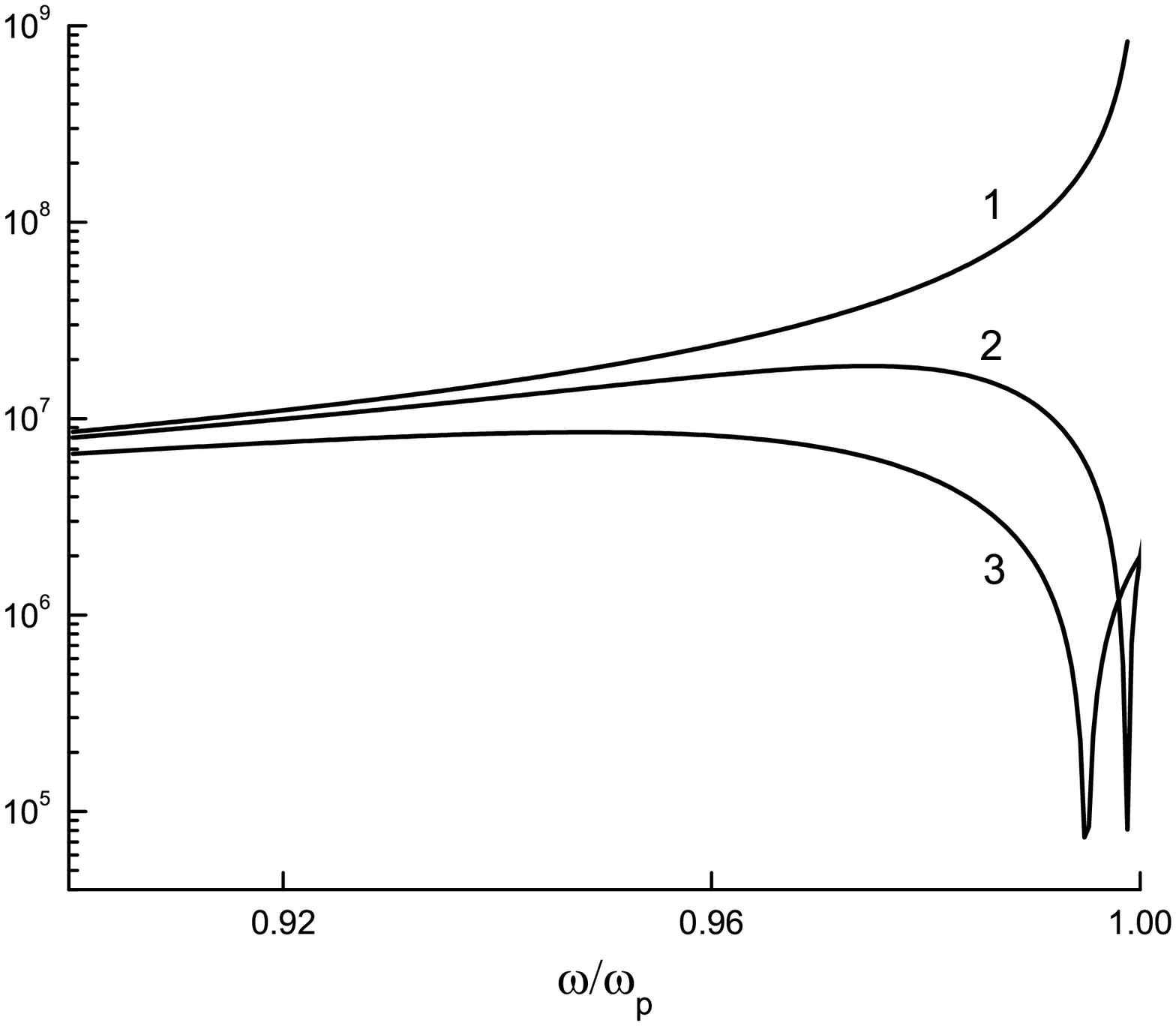}
\noindent\caption{Real part of wave number,
curves $1,2,3$ correspond to values of electron collision frequencies
$\nu=10^{-5}\omega_p,5\cdot 10^{-2}\omega_p,10^{-1}\omega_p$ 1/sec.
Film thickness $d=10$ nm, coefficient of specular reflection
$p=1$ (case of specular reflection).
}\label{rateIII}
\end{figure}

\begin{figure}[h]\center
\includegraphics[width=16.0cm, height=8cm]{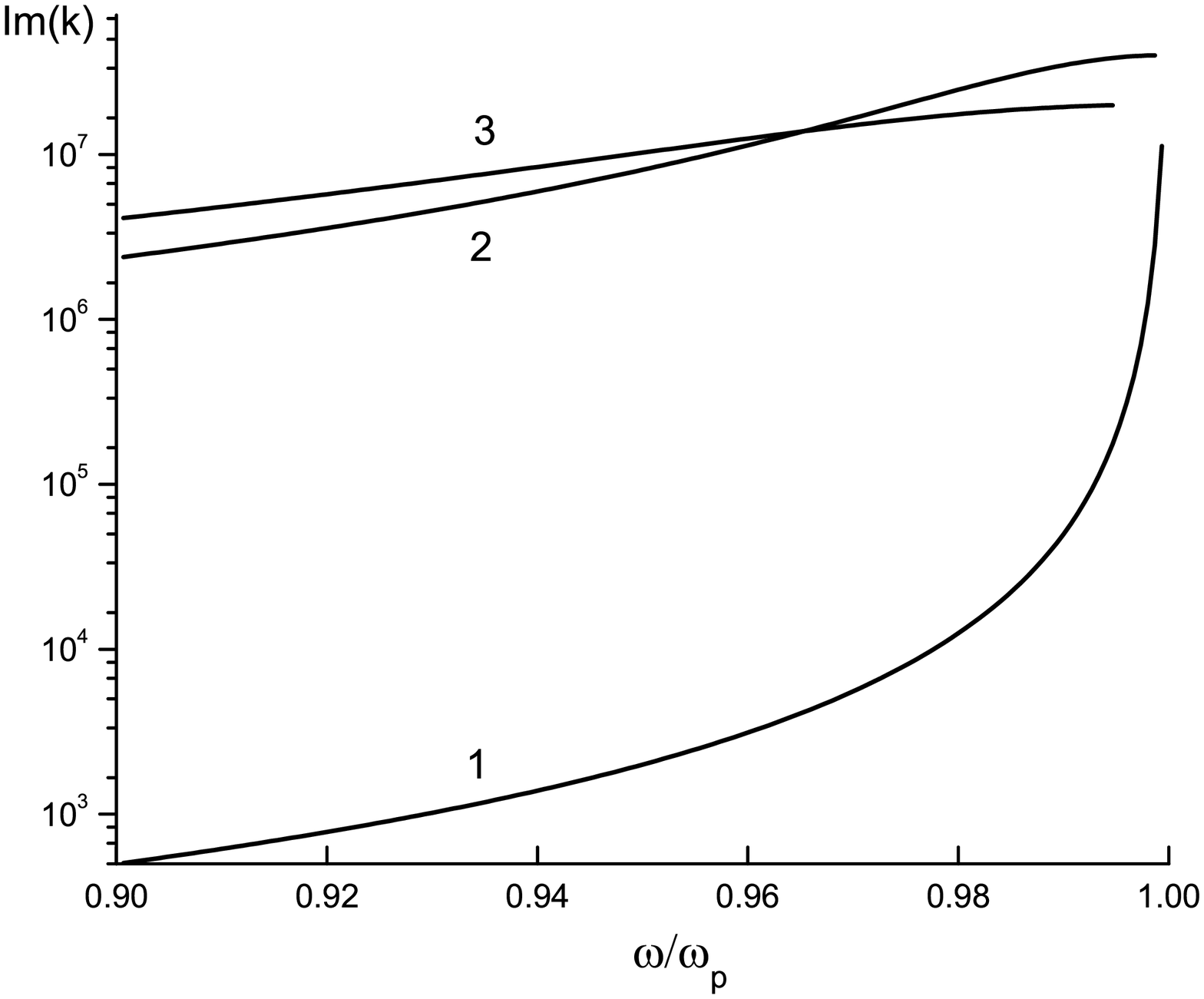}
\noindent\caption{Imaginary part of wave number,
curves $1,2,3$ correspond to values of electron collision frequencies
$\nu=10^{-5}\omega_p,5\cdot 10^{-2}\omega_p,10^{-1}\omega_p$ 1/sec.
Film thickness $d=10$ nm, coefficient of specular reflection
$p=1$ (case of specular reflection).
}\label{rateIII}
\includegraphics[width=16.0cm, height=8cm]{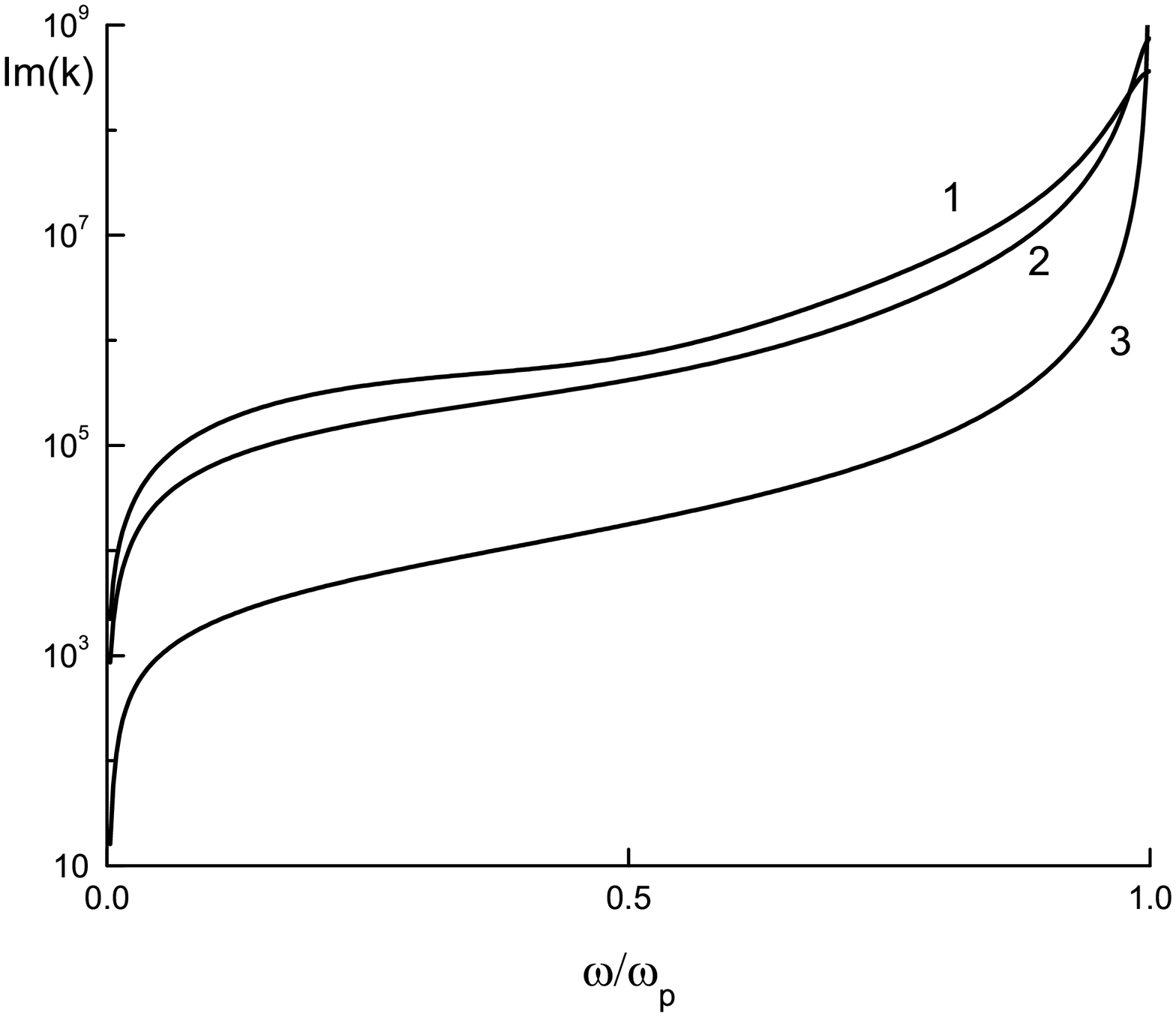}
\noindent\caption{Imaginary part of the wave number $\Im k(\omega)$.
Curves $1,2,3$ correspond to values of coefficient of specular
reflection $p=0,0.5,1$. Electron collision frequencies is equal
$\nu=10^{-3}\omega_p$ 1/sec. Film thickness $d=1$ nm.
}\label{rateIII}
\end{figure}


\begin{thebibliography}{99}


\bibitem{F66}{\it Fuchs R., Kliewer K.L., Pardee W.J.}
Optical properties of an ionic crystal slab
// Phys. Rev. 1966. V. 150. No. 2. P. 589--596.

\bibitem{F69}{\it Jones W.E., Kliewer K.L., Fuchs R.}
Nonlocal theory of the optical properties of thin metallic films
// Phys. Rev. 1969. Vol. 178. No. 3. P.  1201--1203.

\bibitem{F69-2}{\it Kliewer K.L., Fuchs R.}
Optical propertues of an electron gas: Further studies of a nonlocal
description// Phys. Rev. 1969.  Vol. 185. No. 3. P. 805 -- 913.

\bibitem{K} {\it Kondratenko A.N.} Penetration of waves in
plasma. M: Atomizdat, 1979. 232 P. (in russian).

\bibitem{A2004}{\it Antonets I.V., Kotov L.N., Nekipelov S.V.,
Karpushov E.N.}
Conducting and reflecting properties of thin metal films//
Technical Physics, vol. 49, issue 11, pp. 1496-1500.

\bibitem{F09}
{\it Paredes-Ju\'{a}rez A.,  D\'{i}as-Monge F., Makarov N.M.,
P\'{e}res-Rodr\'{i}gues F.} Nonlocal effects
in the electrodynamics of metallic slabs. JETP Lett, 90:9
(2010), 623--627.


\bibitem{Economou}{\it Economou E. N. }
Surface Plasmons in Thin Films //
Phys. Rev. 1969. V. 182. No. 2. P. 539–554.


\bibitem{S}{\it Sondheimer E. H.} The mean free path of electrons in
metals // Advances in Physics, 2001, v. 50, No. 6, 499--537.


\bibitem{R1988}{\it  Raether H.}
Surface Plasmons on Smooth and Rough
Surfaces and on Gratings. Springer, Berlin, 1988. P. 133.
Published 1988 by Springer-Verlag in Berlin, New York .

\bibitem{OPT2003}{\it  Chang R.,  Chiang H.P.,  Leung P.T., Tse W.S.}
Nonlocal electrodynamic effects in the optical excitation
of the surface plasmon resonance
// Optics Communications. V. 225. 2003. P. 353–-361.

\bibitem{JPh2008}
{\it  Maaroof A.I.,  Gentle A.,  Smith G.B.,  Cortie M.B.}
Bulk and surface plasmons in highly nanoporous gold films
// J. Phys. D: Appl. Phys. V. 40. 2007. P. 5675–-5682.


\bibitem{RPP2007}{\it  Pitarke J.M.,  Silkin V.M.,  Chulkov E.V. and
Echenique P.M.}
Theory of surface plasmons and surface-plasmon polaritons
// Rep. Prog. Phys. 2007.  Vol. 70. P. 1–87.


\bibitem{JPh2008}{\it  Anttu N.,  Xu H.Q.}
Light scattering and plasmon resonances in a metal
film with sub-wavelength nano-holes
// J. Phys.: Conference Series.  \,V. 100. 2008. P. 1-4. 052037

\bibitem{Arx2009-1}{\it  Cade N. I., Ritman-Meer C. T.,  Richards  D.}
Strong coupling of localized plasmons and molecular
excitons in nanostructured silver films
// arXiv:0904.2674v1 [cond-mat.mes-hall]. 2009. 4 P.

\bibitem{Arx2009-2}{\it  Apostol M.,  Vaman G.}
Plasmons and polaritons in a semi-infinite plasma and a plasma slab
// arXiv:0904.2662v1 [physics.optics]. 2009. 20 P.

\bibitem{LY2008}
{\it Latyshev A.V., Yushkanov A.A.}
Plasma in a Metal Layer Exposed  to an RF Electric Field//
Technical Physics, 2008, Vol. 53, No. 5, pp. 562--570.

\bibitem{Ina}{\it Inagaki T., Motosuga M., Arakawa E.T. and Goudonnet J.P.}
Coupled surface plasmons in periodically corrugated thin silver films//
Phys. Rev. B. 1985. Vol. 32. Nu. 10. P. 6238--6245.


\bibitem{LYarx2010} {\it Latyshev A.V., Yushkanov A.A.}
Surface Plasmons in Thin Metallic Films//arXiv:1010.2060
[math-ph math.MP physics.optics phy\-sics.plasm-ph],  11 Oct 2010.

\bibitem{Landau8}
{\it Landau L.D., Lifshits E.M.}
Electrodynamics of Continuous Media,
Butterworth-Heinemann (Jan 1984). P. 460.



\bibitem{Landau10}
{\it Lifshits E.M., Pitaevskii L.P.}
Physical Kinetics,
Butterworth-Heinemann (Jan 1981). P. 625.

\end{thebibliography}
\end{document}